\documentclass[conference]{IEEEtran}

\usepackage{graphicx}
\usepackage{datatool}
\usepackage{booktabs} 
\usepackage{cite}
\usepackage[most]{tcolorbox}
\definecolor{bg}{RGB}{247,247,247}

\def\BibTeX{{\rm B\kern-.05em{\sc i\kern-.025em b}\kern-.08em
    T\kern-.1667em\lower.7ex\hbox{E}\kern-.125emX}}

\begin{document}

\title{Identifying Information Technology Research Trends through Text Mining of NSF Awards}

\author{

\IEEEauthorblockN{Said Varlioglu, Hazem Said, Murat Ozer, Nelly Elsayed}
\IEEEauthorblockA{
\textit{School of Information Technology}\\
\textit{University of Cincinnati}\\
Cincinnati, OH, USA \\
varlioms@mail.uc.edu, saidhm@ucmail.uc.edu, ozermm@ucmail.uc.edu, elsayeny@ucmail.uc.edu}
}



\maketitle
\pagestyle{plain}

\begin{abstract}
Information Technology (IT) is recognized as an independent and unique research field. However, there has been ambiguity and difficulty in identifying and differentiating IT research from other close variations. Given this context, this paper aimed to explore the roots of the Information Technology (IT) research domain by conducting a large-scale text mining analysis of 50,780 abstracts from awarded NSF CISE grants from 1985 to 2024. We categorized the awards based on their program content, labeling human-centric programs as IT research programs and infrastructure-centric programs as other research programs based on the IT definitions in the literature. This novel approach helped us identify the core concepts of IT research and compare the similarities and differences between IT research and other research areas. The results showed that IT differentiates itself from other close variations by focusing more on the needs of users, organizations, and societies.


\end{abstract}

%
\IEEEpeerreviewmaketitle

\section{Introduction and Background}

The term information technology (IT) was first used in the late 1950s in an article published in the Harvard Business Review \cite{leavitt1958management}. Coining the term and proposing an information theory are important steps in solving business management needs at that time \cite{shannon1948mathematical}.
This early adaptation of the term suggests that IT has begun to emerge as a discipline predating computer science based on the need for information management in the business sector. Therefore, limiting "IT" to a business term was necessary during that period. Over the years, IT has evolved to encompass all aspects of life, including personal life, academia, and business management~\cite{rowe2011assessment}.

From a historical perspective, Computer Science (CS) became a research discipline in 1968\cite{austing1979curriculum}, Information Systems (IS) was defined as a distinct area of computing in 1981 \cite{adams1982dpma}, and Information Technology undergraduate college programs began to emerge in the late 1990s \cite{helps2010evolving}. The IT discipline was officially recognized with the publication of the first ACM/IEEE-CS curriculum guidelines for IT bachelor's degree programs in 2008 \cite{lunt2008curriculum, Sabin2017}.

After launching Information Technology (IT) programs in universities \cite{helps2010evolving}, they have maintained their attractiveness in academia \cite{rowe2011assessment}. IT research has been an emerging domain \cite{pham2005constituting}, recognizing an independent and unique research field different from Computer Science and related fields such as Information Systems \cite{reichgelt2004comparison}.

Although academia coined the IT term earlier than Computer Science and Information Systems, there is still concept/term confusion on the definition and scope of IT \cite{woratschek2009defining}. Part of the confusion is that the computerization era has invaded the borders of Information Technology by compromising information in the form of the digital format \cite{lenox2003too},\cite{onn2013mini}. For example, Lenox conducted a survey \cite{lenox2003too} to examine the differences or similarities between IT and IS disciplines among diverse faculty members and students. The survey results revealed that 62\% of respondents stated "Information Systems (IS) includes Information Technology (IT)," whereas 34\% of respondents stated "Information Technology (IT) includes Information Systems (IS)" \cite{lenox2003too}. 


On the other hand, National Research Council \cite{national2000making} adopts a broader definition and defines IT as all computational sciences, including computer and information science. Similarly, World Bank uses IT term in a broad manner by addressing IT and Informatics as interchangeable concepts \cite{hanna1993information}. 

Ekstrom et al. \cite{ekstrom2006information} go one step further and define IT as "meeting the needs of users in an organizational and societal context through the selection, creation, application, integration, and administration of computing technologies" \cite{ekstrom2006information}. According to the authors, "IT in its broadest sense encompasses all aspects of computing technology." Some scholars conducted research to explore perceptions about the mixed terms.

In 2017, the curriculum guidelines for IT bachelor's degree programs refined the definition as "\textit{IT is the study of systemic approaches to select, develop, apply, integrate, and administer secure computing technologies to enable users to accomplish the personal, organizational, and societal goals} \cite{Sabin2017}.

Hersh's \cite{hersh2009stimulus} approach is different from his counterparts by formalizing informatics as "\textit{Informatics = People + Information + Technology}." This was one of the first attempts to include the "\textit{People}" term in a definition. In 2021, Said et al. \cite{said2021framework} presented a novel framework aimed at defining IT by incorporating the "\textit{Need}" and "\textit{Solution}" keywords. They proposed that the four components are the key elements of \textit{IT: "People," "Technology," "Information," "Needs/Solutions"}. 

In recent research, the definition of Information Technology has evolved to a more mature form, integrating key concepts of the discipline. Within this context, IT is defined as "the study of solutions and needs that connect people, information, and the technology of the time" \cite{jayathilake2024impact}.

The examples above demonstrate that many researchers have been seeking an answer to the scope of each discipline over the last three decades. Examinations of university programs \cite{georgemason}, \cite{ucsoit} can also lead to confusion because universities define the terms differently to position themselves for the ongoing needs of students. 


This paper aims to explore the roots of the Information Technology (IT) research domain by conducting a large-scale text mining analysis of 50,780 abstracts from awarded NSF CISE grants between 1985 and 2024. We categorized the awards based on their program content, labeling human-centric programs as "IT Research" and infrastructure-centric programs as "Other Research". This classification is derived from the definitions proposed by Said et al. \cite{said2021framework} and the recent definitions, which emphasize the integration of people, information, and technology to address needs and provide solutions \cite{jayathilake2024impact}. This approach helped us identify core concepts of IT research and compare the similarities and differences between IT research and other research areas.

\section{Method}

\subsection{Data}

The current study utilized NSF-CISE awards data, which included 50,780 research abstracts between 1985 and 2024. We analyzed data on the variables of \textit{year, award amount, institution name, institution state, abstract} and \textit{program name}. To determine the abstracts of IT research and its close variations, we used the \textit{program} variable, which helps to roughly understand the scope and the field of each award.

\subsection{Inclusion Criteria}

Using Ekstrom et al.'s \cite{ekstrom2006information}, and Reichgelt et al. are \cite{reichgelt2004comparison}, \cite{reichgelt2014information} inclusive definition, we filtered the abstracts to label them as "IT Research" using the specific terms below. These terms are used under human-centric and society-related programs such as \textit{Information Technology Research, Human-Computer Interaction, Digital Society, Digital Government, Advanced Learning Technologies}, etc.

\begin{tcolorbox}[colback=bg,boxrule=-4pt,arc=0pt]
\begin{itemize}
\scriptsize 
    \item human
    \item society
    \item csforall
    \item ethics \& values
    \item science of design
    \item science of science
    \item digital government
    \item science of learning
    \item information integration
    \item information \& knowledge
    \item smart and connected health
    \item education and human resources
    \item advanced learning technologies
\end{itemize}
\end{tcolorbox}

\begin{figure}[ht]
    \centering
    \includegraphics[width=\linewidth]{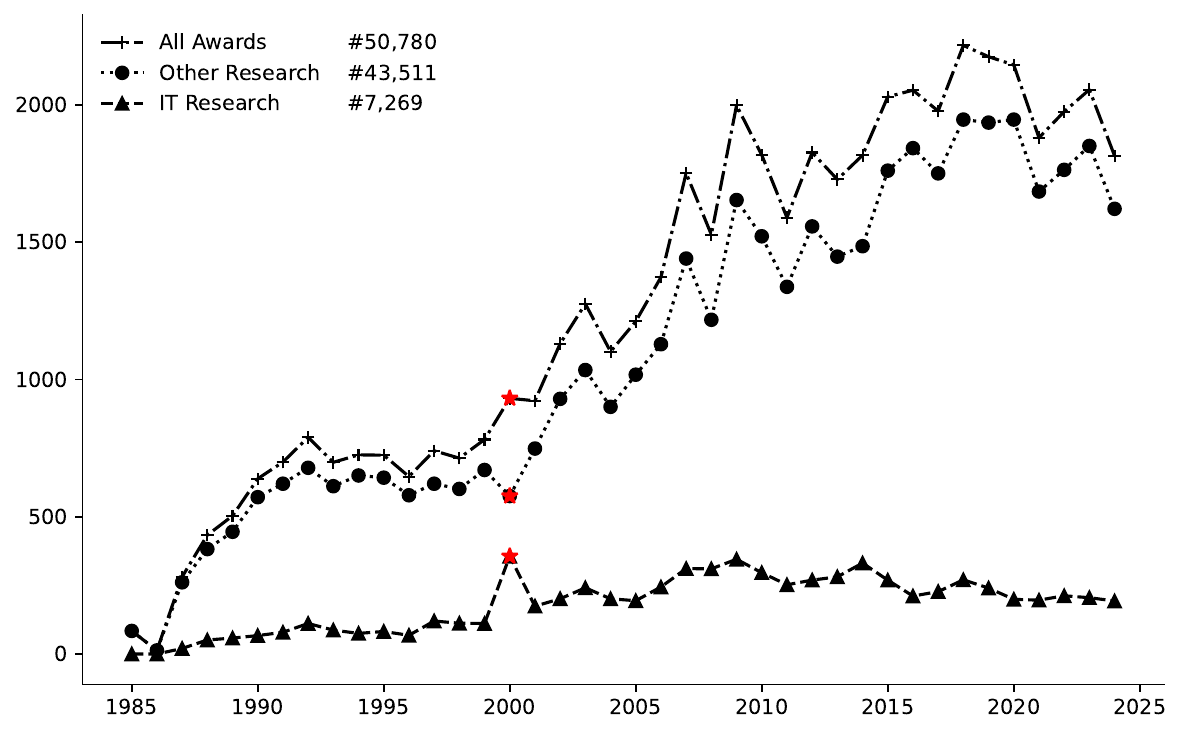}
    \caption{NSF CISE award trends from 1985–2024.}
    \label{nsf_trend}
\end{figure}

\begin{figure}[ht]
    \centering
    \includegraphics[width=\linewidth]{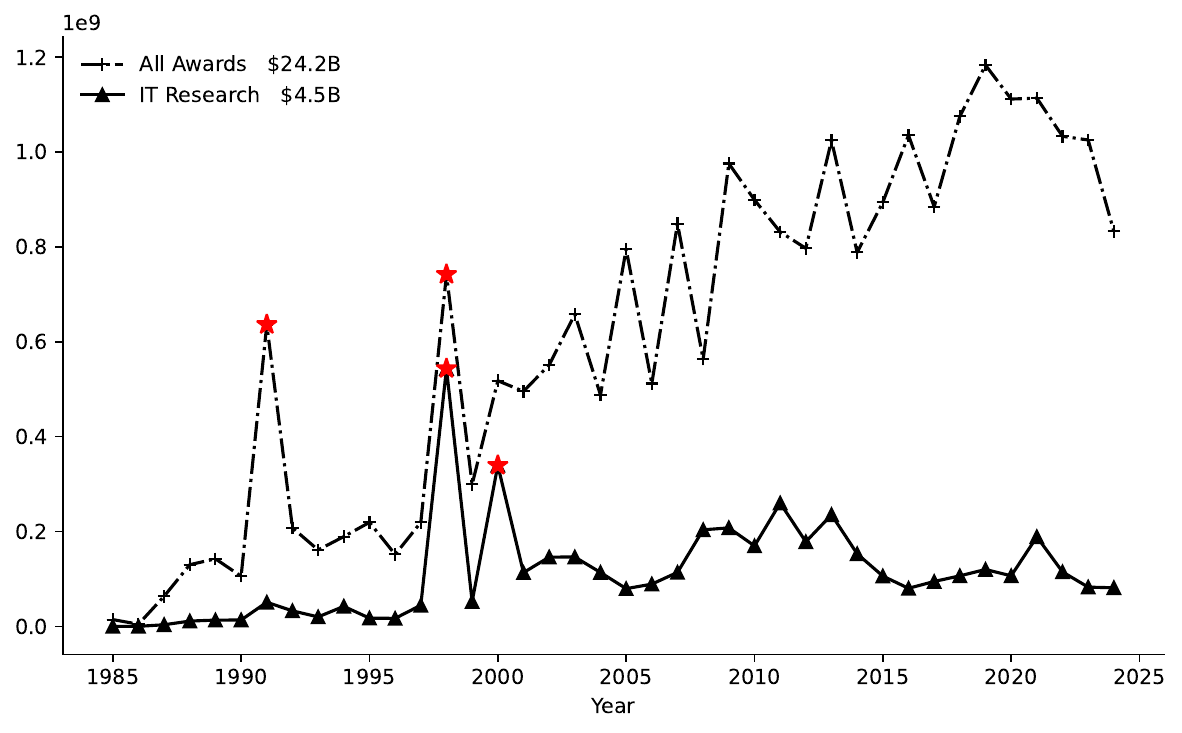}
    \caption{NSF CISE award funding trends from 1985–2024.}
    \label{IT_trend}
\end{figure}

\subsection{Exclusion Criteria}

We excluded the other research programs such as \textit{Computer Systems Architecture, Network Infrastructure, Algorithms, Cyber-Physical Systems} and labeled them as "Other Research" programs in order to make the distinction as dictated in Ekstrom et al. \cite{ekstrom2006information} criteria for non-IT-focused work.

\subsection{Analytical Techniques}

\begin{table*}[h!]
    \centering
    \caption{Top 10 States by IT Research Funding with Comparison to Other Research Areas (1985--2024)}
    \begin{tabular}{lc|cc|cc}
        \hline
        \textbf{State} & \textbf{Total Funding} & \textbf{Other Research} & \textbf{Other Research \%} & \textbf{IT Research} & \textbf{IT Research \%} \\
        \hline
        California & \$3.9B & \$3.0B & 76.9\% & \$903M & 23.1\% \\
        Illinois & \$2.6B & \$2.0B & 77.4\% & \$588M & 22.6\% \\
        Massachusetts & \$1.7B & \$1.3B & 76.9\% & \$397M & 23.1\% \\
        Pennsylvania & \$1.8B & \$1.5B & 80.0\% & \$366M & 20.0\% \\
        New York & \$1.7B & \$1.5B & 84.1\% & \$276M & 15.9\% \\
        \textbf{Texas} & \textbf{\$1.8B} & \textbf{\$1.6B} & \textbf{90.3\%} & \textbf{\$171M} & \textbf{9.7\%} \\
        Michigan & \$645M & \$512M & 79.3\% & \$133M & 20.7\% \\
        \textbf{District of Columbia} & \textbf{\$394M} & \textbf{\$265M} & \textbf{67.3\%} & \textbf{\$129M} & \textbf{32.7\%} \\
        Maryland & \$571M & \$462M & 80.9\% & \$109M & 19.1\% \\
        Washington & \$441M & \$339M & 76.9\% & \$102M & 23.1\% \\
        \hline
    \end{tabular}
    \label{top10_states_percentages}
\end{table*}

\begin{table*}[h!]
    \centering
    \caption{Top 10 Institutions by IT Research Funding with Comparison to Other Research Areas (1985--2024)}
    \begin{tabular}{llc|cc|cc}
        \hline
        \textbf{Institution} & \textbf{Type} & \textbf{Total Funding} & \textbf{Other Research} & \textbf{Other Research \%} & \textbf{IT Research} & \textbf{IT Research \%} \\
        \hline
        Uni. of Illinois at Urbana-Champaign & Public & \$1.8B & \$1.4B & 74.1\% & \$473M & 25.9\% \\
        \textbf{Uni. of California-San Diego} & Public & \$1.0B & \$711M & \textbf{67.7\%} & \$339M & \textbf{32.3\%} \\
        Carnegie-Mellon University & Private & \$819M & \$639M & 78.0\% & \$180M & 22.0\% \\
        Massachusetts Institute of Technology & Private & \$507M & \$376M & 74.0\% & \$132M & 26.0\% \\
        Uni. of California-Berkeley & Public & \$426M & \$329M & 77.3\% & \$97M & 22.7\% \\
        Uni. of Washington & Public & \$380M & \$285M & 74.8\% & \$96M & 25.2\% \\
        Uni. of Michigan - Ann Arbor & Public & \$361M & \$266M & 73.7\% & \$95M & 26.3\% \\
        \textbf{Computing Research Association} & Private & \$159M & \$67M & \textbf{42.2\%} & \$92M & \textbf{57.8\%} \\
        Uni. of California-Los Angeles & Public & \$321M & \$229M & 71.5\% & \$91M & 28.5\% \\
        \textbf{Cornell University} & Private & \$479M & \$390M & \textbf{81.4\%} & \$89M & \textbf{18.6\%} \\
        \hline
    \end{tabular}
    \label{top10_institutions_percentages}
\end{table*}

We used Natural Language Toolkit (NLTK) and Statistical Natural Language Processing (NLP) libraries written in the Python programming language \cite{bird2009natural}. The dataset was examined for missing abstracts to remove null rows. The raw abstract texts were cleaned, converting them to lowercase, stripping newline characters, and removing the punctuation marks.

To conduct text mining, the abstracts were tokenized into individual words, excluding English stopwords (e.g., “the,” “is,” “and”) using NLTK’s predefined stopword list. 

To normalize word forms and reduce morphological variations, lemmatization was applied using NLTK’s WordNet lemmatizer; each word token was reduced to its base or dictionary form (lemma). For example, words such as “technologies” or “users” were normalized to “technology” and “user.”

\section{Comparative Analysis of Funding Distributions}

The categorization approach labeled 7,269 awards as "IT Research" and 43,511 as "Other Research" out of 50,780 abstracts. Figure \ref{nsf_trend} shows the trend of NSF awards from 1985 to 2024. Notably, in the year 2000, while IT Research increased, Other Research experienced a relative decline. However, this was not repeated in the other periods. Thus, we provide the comparative analysis details in the following subsection.

As displayed in Table \ref{top10_states_percentages}, the District of Columbia (D.C.) dedicated 32.7\% of its total funding to IT research. This indicates D.C., as the center of the US government, focuses on human-centric and societal programs that are prevalent in federal departments.


The coastal states that are California, Massachusetts, and Washington have a balanced approach. They allocate around 23\% of their funding to IT research, suggesting that while these states hold large-scale research portfolios, they still keep their commitment to human-centric IT research domains.

Table \ref{top10_institutions_percentages} compares the distribution of IT research funding between public and private institutions. Public universities received a higher amount of IT research funding (\$1.26 billion) with a larger proportion (24.8\%) to human-centric IT research compared with private institutions (20.9\%).

As a remarkable contribution among the institutions, the Computing Research Association (CRA) allocated 57.8\% of its funding to the IT research domains. In contrast, Cornell University, despite receiving massive research funding (\$479 million), allocates only 18.6\% to IT research, contributing more to other infrastructure-centric research domains.

\section{Comparative Analysis of Word Distributions}

Analyzing 7,269 IT Research and 43,511 Other Research abstracts showed that IT Research tends to have longer abstracts on average (mean = 198 tokens) compared to Other Research (mean = 181 tokens).

To explore the most frequent terms used in IT Research and Other Research abstracts, we showed the top 50 most frequent terms in each category as seen in Table \ref{final_top50_terms}.

\begin{table}[ht]
\centering
\caption{Top 50 Most Frequent Terms (Excluding General Terms) in IT Research and Other Research Abstracts}
\begin{tabular}{c|lc|lc}
\hline
\textbf{N} & \textbf{IT Research Term} & \textbf{\%} & \textbf{Other Research Term} & \textbf{\%} \\
\hline
1 & student & 0.67\% & network & 0.72\% \\
2 & design & 0.63\% & application & 0.56\% \\
3 & technology & 0.51\% & student & 0.53\% \\
4 & support & 0.49\% & model & 0.51\% \\
5 & model & 0.47\% & design & 0.49\% \\
6 & work & 0.45\% & algorithm & 0.48\% \\
7 & user & 0.44\% & support & 0.47\% \\
8 & application & 0.40\% & problem & 0.43\% \\
9 & science & 0.40\% & science & 0.41\% \\
10 & develop & 0.38\% & software & 0.40\% \\
11 & computer & 0.36\% & learning & 0.39\% \\
12 & network & 0.35\% & computing & 0.38\% \\
13 & tool & 0.35\% & computer & 0.37\% \\
14 & development & 0.34\% & technique & 0.36\% \\
15 & community & 0.33\% & develop & 0.33\% \\
16 & learning & 0.32\% & method & 0.33\% \\
17 & \textbf{human} & 0.31\% & technology & 0.32\% \\
18 & researcher & 0.31\% & development & 0.31\% \\
19 & \textbf{interaction} & 0.31\% & tool & 0.30\% \\
20 & approach & 0.30\% & \textbf{performance} & 0.30\% \\
21 & method & 0.30\% & approach & 0.30\% \\
22 & \textbf{social} & 0.30\% & program & 0.30\% \\
23 & provide & 0.30\% & community & 0.29\% \\
24 & software & 0.28\% & analysis & 0.29\% \\
25 & algorithm & 0.28\% & work & 0.28\% \\
26 & technique & 0.28\% & user & 0.26\% \\
27 & problem & 0.28\% & broader & 0.26\% \\
28 & study & 0.28\% & provide & 0.25\% \\
29 & goal & 0.26\% & \textbf{foundation} & 0.25\% \\
30 & environment & 0.26\% & computational & 0.24\% \\
31 & computing & 0.26\% & evaluation & 0.24\% \\
32 & analysis & 0.24\% & researcher & 0.23\% \\
33 & \textbf{field} & 0.24\% & goal & 0.23\% \\
34 & \textbf{team} & 0.23\% & communication & 0.23\% \\
35 & \textbf{conference} & 0.23\% & \textbf{resource} & 0.23\% \\
36 & \textbf{understanding} & 0.23\% & \textbf{security} & 0.22\% \\
37 & \textbf{people} & 0.22\% & \textbf{university} & 0.22\% \\
38 & program & 0.22\% & study & 0.21\% \\
39 & \textbf{group} & 0.22\% & including & 0.21\% \\
40 & computational & 0.21\% & \textbf{intellectual} & 0.21\% \\
41 & \textbf{activity} & 0.21\% & \textbf{merit} & 0.20\% \\
42 & evaluation & 0.21\% & \textbf{large} & 0.20\% \\
43 & \textbf{interface} & 0.20\% & \textbf{framework} & 0.20\% \\
44 & \textbf{process} & 0.20\% & environment & 0.20\% \\
45 & communication & 0.20\% & \textbf{proposed} & 0.20\% \\
46 & broader & 0.20\% & \textbf{challenge} & 0.20\% \\
47 & including & 0.20\% & developed & 0.20\% \\
48 & developed & 0.20\% & \textbf{machine} & 0.19\% \\
49 & \textbf{workshop} & 0.19\% & \textbf{mission} & 0.19\% \\
50 & control & 0.19\% & control & 0.19\% \\
\hline
\end{tabular}
\label{final_top50_terms}
\end{table}

While exploring the most frequent terms, we removed high-frequency but semantically broad terms commonly used in both research categories, such as \textit{research}, \textit{project}, \textit{system}, \textit{data}, \textit{impact}, \textit{use}, \textit{using}, \textit{also}, \textit{new}, \textit{result}, \textit{used}, \textit{area}, \textit{many}, \textit{nsf}, \textit{pi}, \textit{well}, \textit{award}, and \textit{based}.

To identify the unique terms in one group but absent in the other, we created Table~\ref{unique_terms_sorted}. It indicates that IT Research features human and society related terms, such as \textit{human}, \textit{social}, \textit{interaction}, \textit{people}, and \textit{understanding}. These terms represent the primary domains of IT research, which are key concepts in the human-centered computing discipline.

In contrast, the unique terms in Other Research were dominated by infrastructure-centric language such as \textit{security}, \textit{machine}, \textit{framework}, \textit{resource}, and \textit{performance}. These terms indicate the priorities in foundational engineering. 


\begin{table}[ht]
\centering
\caption{Distinctive Terms Unique to Top 50 in IT Research and Other Research Abstracts}
\begin{tabular}{cll}
\hline
\textbf{N} & \textbf{Unique to IT Research} & \textbf{Unique to Other Research} \\
\hline
1  & human          & machine       \\
2  & social         & performance        \\
3  & interaction    & security      \\
4  & people         & resource       \\
5  & understanding  & framework    \\
6  & activity       & foundation     \\
7  & conference     & intellectual   \\
8  & team           & challenge      \\
9  & group          & university     \\
10 & interface      & mission        \\
11 & process        & merit          \\
12 & field          & proposed       \\
13 & workshop       & large          \\
\hline
\end{tabular}
\label{unique_terms_sorted}
\end{table}

\begin{table}[h]
  \caption{Usage Rate of Common Words per Abstract Across Research Categories}
\begin{center}
\begin{tabular}{l l l l}
\hline
\textbf{Word} & \textbf{Rate per IT Abs.} & \textbf{Rate per Other Abs.} & \textbf{Diff.} \\
\hline
information  & \textbf{0.67} & 0.36 & \textbf{+0.31} \\
user         & \textbf{0.43} & 0.20 & \textbf{+0.23} \\
human        & \textbf{0.27} & 0.11 & \textbf{+0.16} \\
interaction  & \textbf{0.26} & 0.09 & \textbf{+0.16} \\
social       & \textbf{0.21} & 0.07 & \textbf{+0.15} \\
society      & \textbf{0.13} & 0.03 & \textbf{+0.10} \\
people       & \textbf{0.14} & 0.06 & \textbf{+0.08} \\
\hline
energy       & 0.07 & \textbf{0.12} & -0.05 \\
processing   & 0.25 & \textbf{0.38} & -0.13 \\
theory       & 0.22 & \textbf{0.35} & -0.13 \\
network      & 0.45 & \textbf{0.62} & -0.17 \\
computing    & 0.44 & \textbf{0.63} & -0.19 \\
performance  & 0.27 & \textbf{0.46} & -0.19 \\
hardware     & 0.10 & \textbf{0.30} & -0.20 \\
\hline
\end{tabular}
\end{center}
\label{tab:word_usage_comparison}
\end{table}

To better understand the specific concepts within each research category, we explored the usage rate of common words per abstract across research categories in Table~\ref{tab:word_usage_comparison}. We calculated the usage rate based on a word appearance in all abstracts in a time period. For example, if the "user" word appeared 100 times in 10,000 words, its usage rate was 1.0\%.

As shown in Table~\ref{tab:word_usage_comparison}, words such as \textit{information}, \textit{user}, and \textit{human} appeared significantly more frequently in IT Research abstracts, indicating a stronger focus on human-centered, user-oriented, and information-driven topics. For instance, the term \textit{information} appeared 1.03 times per IT abstract compared to just 0.61 in Other Research, while \textit{user} and \textit{human} showed similar disparities. In the same context, the term \textit{hardware} appeared 0.10 times per IT abstract compared to 0.30 in Other Research.

To gain more insight into the thematic distinctions between IT Research and Other Research, we explored the top 50 bigrams of each category as shown in Table~\ref{tab:top50_bigrams_final_norownum} and Table~\ref{tab:unique_bigrams}.

The bigrams unique to IT Research focus on user, social, and educational contexts. These include phrases such as \textit{humancomputer interaction}, \textit{social network}, \textit{student participant}, \textit{virtual reality}, and \textit{user interface}, indicating a strong orientation for human interaction, learning environments, and experiential technologies.

In contrast, the bigrams unique to Other Research focus on technical infrastructure, systems optimization, and data-centric innovation with terms such as \textit{data science}, \textit{neural network}, \textit{deep learning}, and \textit{wireless network}.

\begin{table}[ht]
\centering
\caption{Top 50 Most Frequent Bigrams (Excluding Common Terms) in IT Research and Other Research Abstracts}
\begin{tabular}{lc|lc}
\hline
\textbf{IT Research} & \textbf{\%} & \textbf{Other Research} & \textbf{\%} \\
\hline
broader impact             & 0.15\% & broader impact             & 0.20\% \\
intellectual merit         & 0.12\% & intellectual merit         & 0.18\% \\
computer science           & 0.12\% & computer science           & 0.10\% \\
evaluation using           & 0.10\% & evaluation using           & 0.16\% \\
support evaluation         & 0.10\% & support evaluation         & 0.16\% \\
merit broader              & 0.10\% & merit broader              & 0.16\% \\
impact review              & 0.10\% & using foundation           & 0.16\% \\
review criterion           & 0.10\% & review criterion           & 0.16\% \\
award reflects             & 0.10\% & nsf statutory              & 0.16\% \\
reflects nsf              & 0.10\% & statutory mission          & 0.16\% \\
nsf statutory              & 0.10\% & reflects nsf              & 0.16\% \\
statutory mission          & 0.10\% & mission deemed             & 0.16\% \\
mission deemed             & 0.10\% & deemed worthy              & 0.16\% \\
deemed worthy              & 0.10\% & worthy support             & 0.16\% \\
worthy support             & 0.10\% & foundation intellectual    & 0.16\% \\
using foundation           & 0.10\% & impact review              & 0.16\% \\
foundation intellectual    & 0.10\% & award reflects             & 0.16\% \\
machine learning           & 0.07\% & machine learning           & 0.12\% \\
graduate student           & 0.06\% & project develop            & 0.06\% \\
project develop            & 0.06\% & project aim                & 0.06\% \\
research project           & 0.05\% & graduate student           & 0.06\% \\
\textbf{doctoral consortium}        & 0.05\% & research project           & 0.06\% \\
goal project               & 0.05\% & project also               & 0.05\% \\
research community         & 0.05\% & goal project               & 0.05\% \\
\textbf{information technology}     & 0.04\% & \textbf{science engineering}        & 0.05\% \\
\textbf{humancomputer interaction}  & 0.04\% & \textbf{proposed research}          & 0.05\% \\
next generation            & 0.04\% & \textbf{undergrad student}      & 0.04\% \\
high school                & 0.04\% & \textbf{high performance}           & 0.04\% \\
science engineering        & 0.04\% & research community         & 0.04\% \\
proposed research          & 0.04\% & \textbf{undergraduate}     & 0.04\% \\
project also               & 0.04\% & \textbf{big data}                   & 0.04\% \\
\textbf{user interface}             & 0.04\% & \textbf{project develops}           & 0.04\% \\
\textbf{social medium}              & 0.03\% & underrepresented group     & 0.03\% \\
project aim                & 0.03\% & \textbf{data science}               & 0.03\% \\
\textbf{develop new}               & 0.03\% & develop new                & 0.03\% \\
wide range                 & 0.03\% & \textbf{next generation}           & 0.03\% \\
\textbf{research team}                 & 0.03\% & \textbf{wireless network}           & 0.03\% \\
\textbf{united state}             & 0.03\% & \textbf{artificial intelligence}    & 0.03\% \\
\textbf{social network}             & 0.03\% & \textbf{neural network}             & 0.03\% \\
artificial intelligence           & 0.03\% & \textbf{research education}         & 0.03\% \\
\textbf{information system}    & 0.03\% & \textbf{data set}                   & 0.03\% \\
underrepresented group       & 0.03\% & high school                & 0.03\% \\
\textbf{student participant}    & 0.03\% & wide range                 & 0.03\% \\
\textbf{virtual reality}     & 0.03\% & project address            & 0.03\% \\
\textbf{natural language}         & 0.03\% & \textbf{project focus}              & 0.03\% \\
project address        & 0.03\% & \textbf{deep learning}              & 0.03\% \\
\textbf{project outcome}         & 0.03\% & \textbf{programing language}              & 0.03\% \\
\textbf{research develop}         & 0.03\% & \textbf{data analysis}        & 0.03\% \\
\textbf{data mining}         & 0.03\% & computer vision              & 0.03\% \\
computer vision               & 0.03\% & \textbf{sensor network}            & 0.03\% \\
\hline
\end{tabular}
\label{tab:top50_bigrams_final_norownum}
\end{table}

\begin{table}[ht]
\centering
\caption{Bigrams Unique to the Top 50 in IT Research and Other Research}
\begin{tabular}{l|l}
\hline
\textbf{Unique to IT Research}        & \textbf{Unique to Other Research}       \\
\hline
humancomputer interaction             & high performance                        \\
user interface                        & wireless network                        \\
social network                        & sensor network                          \\
social medium                         & neural network                          \\
student participant                   & programming language                    \\
doctoral consortium                   & deep learning                           \\
virtual reality                       & big data                                \\
research team                         & data science                            \\
research develop                      & data analysis                           \\
project outcome                       & data set                                \\
united state                          & project develops                        \\
information system                    & project focus                           \\
information technology                & research education                      \\
natural language                      & undergraduate student                   \\
data mining                           & undergraduate graduate                  \\
\hline
\end{tabular}
\label{tab:unique_bigrams}
\end{table}


\section{Comparing Usage Rate of IT Research Words Time Periods}

\begin{table*}[ht]
\centering
\caption{Top 50 Words by Period in IT Research and Other Research}
\label{MergedTop50Words}
\begin{tabular}{ll|ll|ll|ll}
\hline
\multicolumn{2}{l|}{\textbf{1985--1995}} & \multicolumn{2}{l|}{\textbf{1996--2005}} & \multicolumn{2}{l|}{\textbf{2006--2015}} & \multicolumn{2}{l}{\textbf{2016--2024}} \\
\textbf{IT} & \textbf{Other} & \textbf{IT} & \textbf{Other} & \textbf{IT} & \textbf{Other} & \textbf{IT} & \textbf{Other} \\
\hline
database & algorithm & information & network & student & network & student & support \\
information & problem & application & application & design & student & support & learning \\
model & network & model & problem & technology & application & design & network \\
application & computer & technology & algorithm & information & design & model & student \\
query & design & design & design & work & model & technology & model \\
user & parallel & user & model & network & software & work & application \\
computer & program & technique & information & science & algorithm & science & science \\
design & method & network & technique & application & problem & learning & algorithm \\
problem & model & development & performance & user & information & user & design \\
language & technique & develop & software & model & science & develop & evaluation \\
work & application & work & computer & computer & computing & evaluation & computing \\
technology & support & problem & method & social & computer & community & mission \\
development & performance & tool & development & community & technique & team & intellectual \\
technique & development & database & technology & support & technology & intellectual & merit \\
environment & information & student & work & develop & performance & information & software \\
study & university & algorithm & support & interaction & tool & tool & review \\
support & work & software & student & development & development & mission & criterion \\
method & analysis & computer & analysis & tool & approach & review & reflects \\
processing & software & support & approach & human & analysis & merit & statutory \\
performance & student & provide & develop & software & work & criterion & worthy \\
knowledge & language & method & program & researcher & method & reflects & community \\
theory & science & approach & tool & approach & develop & human & develop \\
approach & communication & analysis & communication & provide & support & statutory & problem \\
process & approach & study & university & problem & program & worthy & technique \\
developed & theory & language & computing & environment & community & computer & method \\
goal & provide & environment & proposed & computing & user & method & technology \\
speech & study & developed & control & field & learning & social & information \\
control & tool & goal & service & learning & computational & researcher & computer \\
provide & develop & query & environment & method & provide & interaction & tool \\
tool & goal & science & provide & algorithm & security & people & approach \\
algorithm & programming & researcher & goal & conference & communication & computing & development \\
management & developed & community & distributed & study & university & provide & user \\
interface & architecture & web & science & technique & researcher & conference & program \\
analysis & high & large & resource & goal & large & development & machine \\
structure & processing & communication & study & computational & resource & goal & researcher \\
develop & computational & process & computational & analysis & study & health & analysis \\
issue & college & performance & wireless & program & goal & study & security \\
object & state & understanding & user & group & proposed & approach & challenge \\
time & time & management & high & control & challenge & algorithm & performance \\
human & access & human & developed & understanding & developed & application & work \\
workshop & efficient & program & architecture & communication & infrastructure & understanding & framework \\
distributed & connection & interface & large & activity & control & activity & provide \\
task & computing & time & internet & people & engineering & group & computational \\
large & environment & distributed & proposal & interface & environment & field & goal \\
group & computation & learning & component & individual & architecture & challenge & resource \\
constraint & distributed & interaction & theory & engineering & high & technique & device \\
network & structure & field & include & health & wireless & help & training \\
researcher & implementation & workshop & time & challenge & education & program & aim \\
agent & control & set & protocol & workshop & framework & network & enable \\
communication & resource & access & efficient & behavior & understanding & analysis & scientific \\
\hline
\end{tabular}
\end{table*}


To find the most commonly used terms in IT research over time, we analyzed word frequency in different time periods. We divided the dataset into four periods: 1985–1995, 1996–2005, 2006–2015, and 2016–2024. For each period, we only included abstracts labeled as \textit{"IT Research"} based on program categories.

We cleaned and combined the lemmatized words from each abstract by removing a set list of general terms (like \textit{system, data, project, use}) and NSF-specific words (like \textit{pi, award, nsf}). We also exclude the filler words that are used in grant writing, such as \textit{one, novel}, and \textit{including}, because they do not add meaningful, domain-specific content.

Table \ref{MergedTop50Words} shows the top 50 most frequent words in the four periods. Words that appeared in only one period's top list and not in any of the others were marked as unique to that period.

Before 1995, most U.S. households had no personal computers, internet, and cell phone. In the 1996-2005 period, most of the U.S. households started to have internet and "dumb" cell phones. In the 2006-2015 period, most U.S. households had fast internet and smartphones~\cite{ws}, \cite{file2013computer}. After 2015, most U.S. households have broadband Internet (high speed such as cable, fiber optic, or cellular data plan for a smartphone) subscription~\cite{ryan2017computer} which contribute to \textit{"individual empowerment, economic growth, and community development"}~\cite{jayakar2016broadband}. Also, after the years of 2014 and 2015, new smart home applications have been launched, and people have started to use smart devices and applications with a rising trend~\cite{miller2015internet}.

\subsection{1985–1995 Period: Emergence of Human-Centric IT Research}

During this period, people started using personal computers, and the Internet began to emerge. Also, graphical user interfaces (GUIs) help people use computers. NSF programs, such as ITR (Information Technology Research), began focusing on projects that explored the human and organizational aspects of computing, including human-computer interaction and digital information systems. These changes helped shape IT as a field that addresses both technical and social aspects of computing.

Between 1985 and 1995, there were 630 abstracts labeled as IT Research and 4,957 under Other Research programs. This decade was a key turning point with the transition to more interactive, user-centered computing approaches.

In this term, IT Research abstracts frequently featured tokens such as:
\textit{database, information, model, application, query, user, computer, design, problem, language}.

Other Research abstracts were dominated by terms like:
\textit{algorithm, problem, network, computer, design, parallel, program, method, model, technique}.

The unique terms found in the IT Research category reflect a growing emphasis on data management, usability, and the societal impact of computing.

In contrast, the Other Research category features heavily infrastructure-oriented language, including terms such as algorithm, network, parallel, and program, reflecting core computer science theory.

\subsection{1996–2005 Period: Growth of Digital Infrastructure and Applied IT}

Between 1996 and 2005, IT research grew rapidly with the rise of open-source software and big investment of the U.S. government. IT expanded significantly across academia, government, and industry. IT research shifted from infrastructure problems to solving complex problems involving people, policy, and knowledge. This led to IT playing a larger role in institutions, education, and society. 

Between 1996 and 2005, a total of 1,778 abstracts were categorized as IT Research, while 7,672 abstracts were categorized as Other Research. This period includes the expansion of the internet, the rise of e-commerce, and the growing integration of information technology into nearly every sector of society.

The most frequent lemme tokens for this period show a continued evolution in research priorities, IT Research top tokens:
\textit{information, application, model, technology, design, user, technique, network, development, develop}. Other Research top tokens:
\textit{network, application, problem, algorithm, design, model, information, technique, performance, software.}

The frequent use of words, information, technology, and models in IT research indicates a growing emphasis on applied technologies and information systems. On the other hand, Other Research maintained its foundation in computational infrastructure, focusing on algorithms, networks, and performance, aligning with the core CS theory.

\subsection{2006–2015 Period: The Rise of Data-Driven and Educational IT}

This period witnessed the rise of social media, smartphones, and significant internet activity. The internet has reshaped both society and science, creating a massive amount of data that should be analyzed and visualized. Cloud platforms like AWS and Google Cloud-enabled scalable research and deployment. During this period, 2,908 abstracts were categorized as \textit{IT Research}, while 14,545 abstracts were categorized under \textit{Other Research} programs.

The top lemme tokens for each research category reflect these major shifts: IT Research top tokens:
\textit{research, project, system, student, data, design, new, technology, information, work.} Other Research top tokens: \textit{project, research, system, data, network, new, student, application, design, model.}

The most frequent lemme tokens for this period show a continued evolution in research priorities, IT Research top tokens:
\textit{student, design, technology, information, work, network, science, application, user, model}. Other Research top tokens:
\textit{network, student, application, design, model, software, algorithm, problem, information, science.}

\subsection{2016–2024 Period: AI, Ethics, and Inclusive IT}

This era has been shaped by artificial intelligence, machine learning, IT ethics, and human-centered design. Also, cyber security appeared as a strong domain. During this period, a total of 1,953 abstracts were categorized as \textit{IT Research}, while 16,337 abstracts fell under \textit{Other Research} programs.

The most frequent lemme tokens for this period show a continued evolution in research priorities, IT Research top tokens:
\textit{student, support, design, model, technology, work, science, learning, user, develop}. Other Research top tokens:
\textit{support, learning, network, student, model, application, science, algorithm, design, evaluation.}

AI-related terms have also emerged strongly, with terms such as intelligence, robot, AI, and machine. In contrast, the Other research category has maintained a focus on systems and performance engineering, such as framework, infrastructure, and computational.

\subsection{Period-Based Comparison of IT and System-Oriented Terms}

\begin{figure*}
    \centering
    \includegraphics[width=1\textwidth]{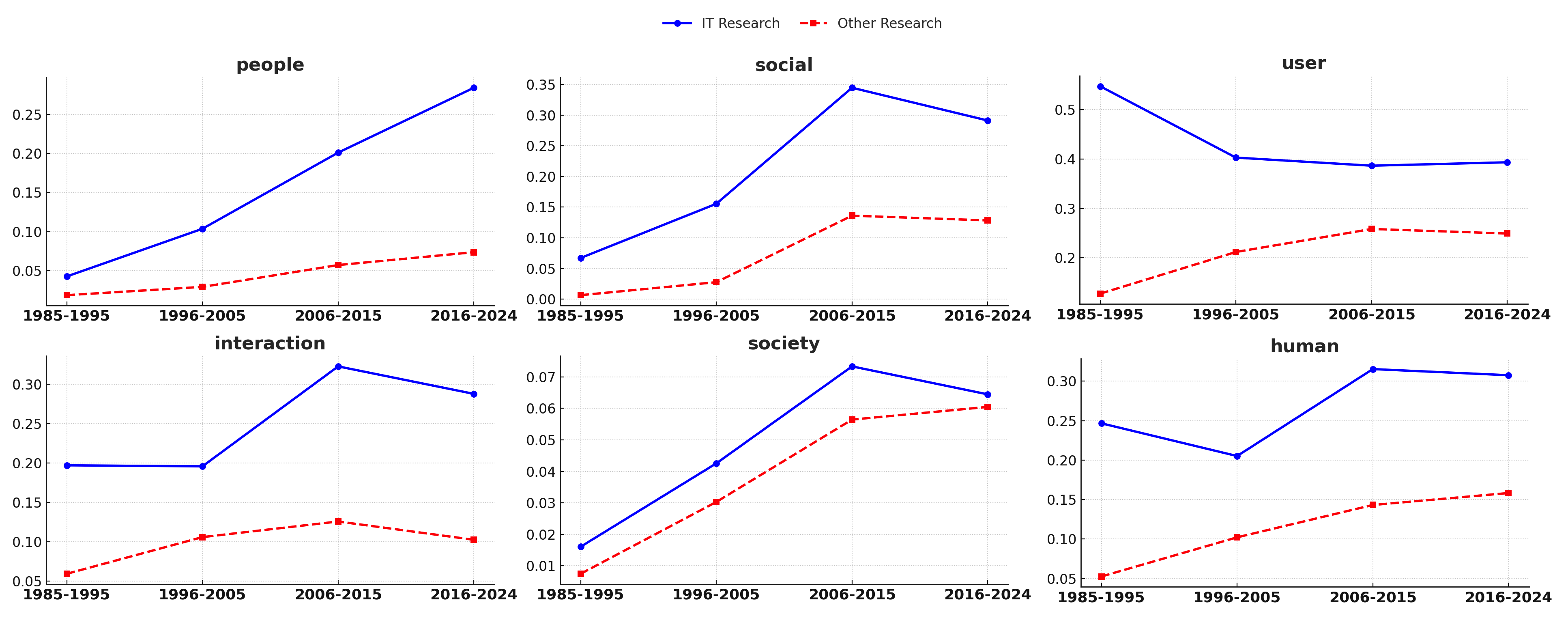}
    \caption{Usage rates of four human-related terms in IT Research abstracts across five periods.}
    \label{human_trend}
\end{figure*}

Figure~\ref{human_trend} shows the usage trends of six human-related words, \textit{people}, social, interaction, society, \textit{human}, and \textit{user}, in IT Research abstracts across five time periods from 1985 to 2024. 

The term \textit{people}, for example, increased from near-zero usage in the earliest period to over 0.25 mentions per abstract in 2021–2024. Similarly, \textit{human}, \textit{society}, and \textit{social} saw growth showing social-oriented research. The term \textit{interaction} is consistently higher in IT research.

\begin{figure*}
    \centering
    \includegraphics[width=1\textwidth]{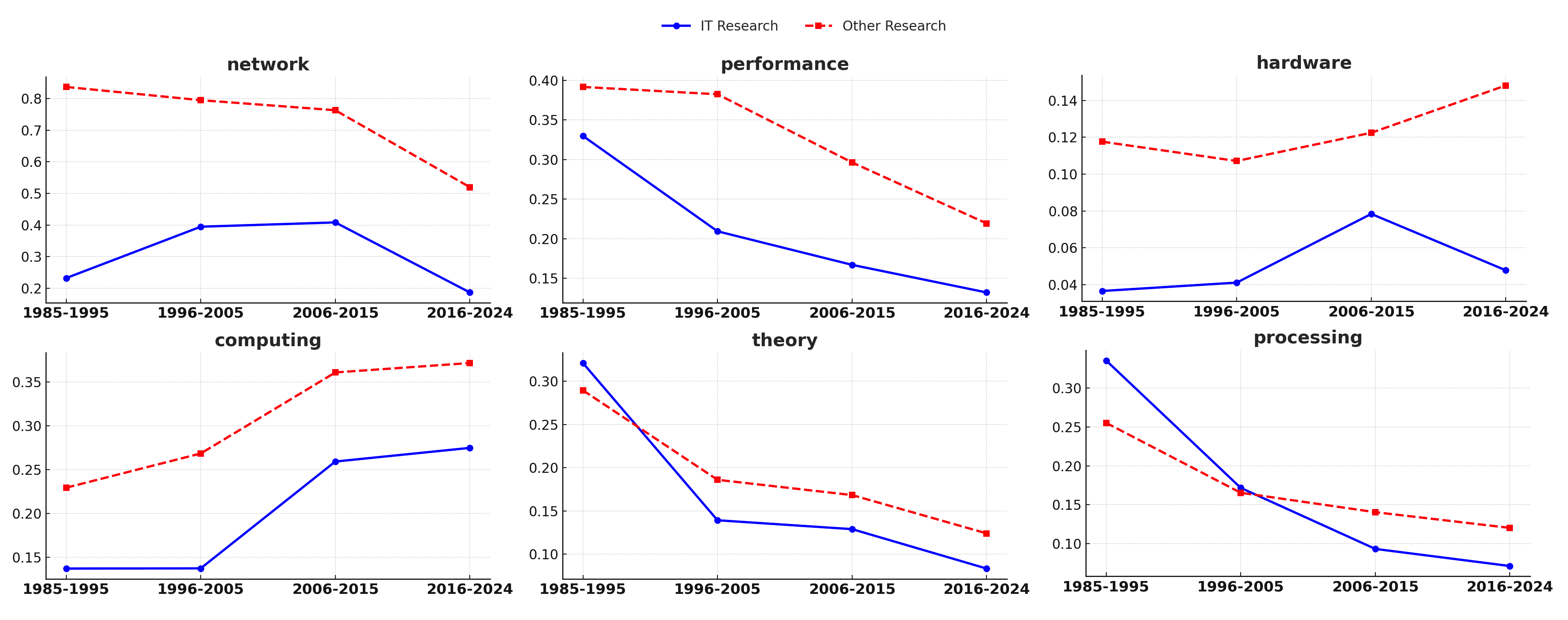}
    \caption{Usage rates of four computing-related terms in IT Research abstracts across five periods.}
    \label{computing_trend}
\end{figure*}

Figure~\ref{computing_trend} shows the usage trends of six computing-related words, \textit{network}, \textit{performance}, \textit{hardware}, \textit{computing}, \textit{theory}, and \textit{processing} across five periods from 1985 to 2024. 

Unlike the steady growth observed in human-related terms, these technical terms \textit{network}, \textit{performance}, \textit{theory}, and \textit{processing} show a decline over time. 

The longitudinal analysis in Figure~\ref{human_trend} shows a critical shift in the Information Technology (IT) research landscape: the growing emphasis on human- and society-centered themes. Specifically, the usage rate of terms like \textit{people}, \textit{social}, and \textit{society} increased more than fivefold. 

\section{Discussion and Conclusion}

In this paper, we aimed to determine the scope of the IT research domain by analyzing historical NSF award abstracts. Analysis revealed that IT research differentiates itself from similar fields by focusing on and emphasizing more human and society-oriented needs and solutions. On the other hand, the research focus of other closely related domains generally stays in more technical and hard science engineering areas. 

In this context, our study aligns with the IT definitions proposed by Said et al. \cite{said2021framework} and recent research, which emphasize the integration of people, information and technology to meet needs and provide solutions \cite{jayathilake2024impact}. The results highlight that the central focus of IT is meeting the needs of users, organizations, and societies using computing technologies. 

The findings validate IT has a distinct focus compared to more technical, infrastructure-centric disciplines, as well as support the user-centered definition of IT research.

This study is one of the first attempts that empirically attempt to identify the scope of information technology. 

Further research can examine more NSF awards and their research papers using the same method or advanced text mining and sentiment analysis methods. Additionally, research publications from various IT schools can be examined to explore how their research differs or is similar to other research disciplines, such as Computer Science and Information Systems.

\bibliographystyle{IEEEtran}
\bibliography{main}
\end{document}